\documentstyle[]{mn}        
\title [Large Scale Structure in a deep IRAS survey]
       {Large Scale Structure in a new deep IRAS galaxy redshift survey} 
\author[S.J. Oliver \et]
       {S.J. Oliver$^1$,
        M. Rowan-Robinson$^1$, 
        T.J. Broadhurst$^2$, 
        R.G. McMahon$^3$, \\
\\
\begin{LARGE}{\rm 
        W. Saunders$^4$, 
        A. Taylor$^5$ 
        A. Lawrence$^5$,
        C.J. Lonsdale$^6$,
        P. Hacking$^6$,
        and
}
\end{LARGE}\\
\\
\begin{LARGE}{\rm 
        T. Conrow$^6$  
}
\end{LARGE}\\
$^1$ Imperial College of Science Technology and Medicine,  Blackett Laboratory,
Prince Consort Road.  London SW7 2BZ.\\
$^2$ Johns Hopkins University, Johns Hopkins Road,  
Laurel, MD 20723-6099, USA.\\
$^3$ Institute of Astronomy, Madingly Road, Cambridge  CB3 0HA.\\
$^4$ Department of Astrophysics, Keeble Road, Oxford OX1 3RH.\\
$^5$ Institute for Astronomy, Blackford Hill, Edinburgh EH9 3HJ.\\
$^6$ IPAC, California Institute of Technology,
770 South Wilson Avenue, Mail Code 100-22, Pasadena, CA 91125, USA.\\
}
\date{Accepted 1995 November 16. Received 1995 November 1; in original
form July 12}

\newcommand{\et}{{\em et al.}}
\newcommand{\kms}{\ifmmode {\rm km\thinspace s}^{-1}
                  \else km\thinspace s$^{-1}$ \fi}      

\newcommand{\var}{{\rm var}}

\def\eps@scaling{.95}
\def\epsscale#1{\gdef\eps@scaling{#1}}
\def\plotone#1{\centering \leavevmode
\epsfxsize=\eps@scaling\columnwidth \epsfbox{#1}}

\newcommand{\mn}{MNRAS}
\newcommand{\apj}{ApJ}
\newcommand{\apjs}{ApJS}
\newcommand{\apjl}{ApJ}
\newcommand{\na}{Nat}

\begin{document}

\maketitle

\begin{abstract}

We present here the first results from two recently completed, fully
sampled redshift surveys comprising 3703 IRAS Faint Source Survey (FSS)
galaxies.  An unbiased counts-in-cells analysis finds a clustering
strength in broad agreement with other recent redshift surveys and at
odds with the standard cold dark matter model.  We combine our data
with those from the QDOT and 1.2~Jy surveys, producing a single
estimate of the IRAS galaxy clustering strength.  We compare the data
with the power spectrum derived from a mixed dark matter universe.
Direct comparison of the clustering strength seen in the IRAS samples
with that seen in the APM-Stromlo survey suggests
$b_O/b_I=1.20\pm0.05$ assuming a linear, scale independent biasing.
We also perform a cell by cell comparison of our FSS-$z$ sample with
galaxies from the first CfA slice, testing the viability of a linear-biasing scheme linking the two.  We are able to rule out models in
which the FSS-$z$ galaxies identically trace the CfA galaxies on
scales 5-20$h^{-1}$Mpc.  On scales of 5 and 10$h^{-1}$Mpc no linear-biasing model can be found relating the two samples.  We argue that
this result is expected since the CfA sample includes more elliptical
galaxies which have different clustering properties from spirals.  On
scales of 20$h^{-1}$Mpc no linear-biasing model with $b_O/b_I<1.70$ is
acceptable. When comparing the FSS-$z$ galaxies to the CfA {\em
spirals}, however, the two populations trace the same structures
within our uncertainties.

\end{abstract}

\begin{keywords}

galaxies: clusters: general --
galaxies: elliptical \& lenticular, cD --
galaxies: spiral --
large-scale structure of Universe --
infrared: galaxies

\end{keywords}
\section{Introduction}\label{intro}

Following the success of IRAS Point Source Catalog (PSC) galaxy redshift
surveys in probing the large scale structure of the local Universe,
which presented serious problems for the standard cold dark matter (CDM)
theory \cite{e90,snature,k91,moore,2jy_lss,fish_pow} two new fainter
redshift surveys have been undertaken to confirm these results at
significantly greater depths.

The first of these surveys (FSS-$z$~I\footnote{These surveys have been
referred to previously as `QCCOD'}) was based on samples drawn from
the IRAS Faint Source Data Base [FSDB, see Moshir
\et~\shortcite{fss}].  On the basis of deep IRAS coverage and freedom
from cirrus contamination, three regions in the Northern Galactic
Hemisphere were selected (these areas are defined in
Table~\ref{areas}). Within these areas we believe the FSDB to be $\sim
99$ per cent complete for $S_{60}\ge 0.2$~Jy (Lonsdale~\et~in
preparation). We thus constructed a catalogue of all FSDB sources
within these areas having good to moderate $60\mu$m fluxes greater
than 0.2~Jy.  (In area E an additional sample was selected with
$0.15\le S_{60}<0.2$ but this is not considered further in this
paper.)  The majority of such sources are galaxies
[e.g. Rowan-Robinson~\et~ \shortcite{rr86},
Lawrence~\et~\shortcite{al86}].  No formal colour-cuts were employed
to exclude either stars or cirrus sources.  The former were excluded
upon inspection of the POSS plates or APM cartoons together with a
case by case examination of the colours.  In the combined FSS samples
only five `stars' lay close to the galaxy colour locus and they had
the same colours as the $60\mu$m excess stars. In addition 13
galaxies were excluded because a bright star lay in the field making
identification and acquisition impossible, these are assumed to be
random line of sight coincidences with little effect on our analysis.
Cirrus sources were rare because of our careful selection of areas (four
sources have been excluded upon examination of maps made from the raw
IRAS data).  This survey was particularly designed to clarify the
evolutionary behaviour of the IRAS population; see contrasting
conclusions of Saunders \et \shortcite{s90} and Fisher
\et~\shortcite{fish_evol}.

The second survey (FSS-$z$~II) was constructed from the Faint Source
Catalog Version 2 \cite{fss}. The area selected traversed the North
Galactic Pole connecting the FSS-$z$~I areas. In other respects the
FSS-$z$~II samples were selected in the same manner as FSS-$z$~I.
Designed principally for large scale structure studies, the flux
completeness in areas P and (to a lesser extent) X is not as good as
other areas and is estimated to be 90 per cent at 0.25 Jy.

The combined surveys cover an area of 1310 deg$^2$ (0.4 sr) and
contain 3728 sources, more than 3600 of which are galaxies.

\begin{table}\centering
\caption{Definition of FSS-$z$ areas.  Areas N, A and E constitute FSS-$z$~I
while P and X constitute FSS-$z$~II.}
\label{areas}
\begin{tabular}{cccr}
    &           \multicolumn{2}{c}{Definition}            & Area/\\
    &                        &                            &sq deg\\ 
 N  & $150\deg\le l\le 210\deg, $ & $ 50\deg\le b\le 70\deg$  & 594\\
 A  & $ 30\deg\le l\le  90\deg, $ & $ 60\deg\le b\le 68\deg$  & 178\\
 E  & $ 70\deg\le l\le  90\deg, $ & $ 50\deg\le b\le 55.5\deg$&  67\\
 P  & $26.5\deg\le\delta\le44.5\deg,$
    & $ b\ge 70\deg$ if $\alpha\le 12^h$ & 476\\
    && or $ b\ge 68\deg$ if $\alpha>12^h$&\\
 X  & $32.5\deg\le\delta\le38.5\deg,$  
    & $b<50\deg$ \& $\alpha\ge8^h$         & 94\\

\end{tabular}                                                          
\end{table}

Redshifts were obtained from the literature for 872 sources; a number
of redshifts were kindly provided in advance of publication by: John
Huchra, Ray Wolstencroft, Quentin Parker \& Roger Clowes and Marc
Davis \& Michael Strauss. Major observation programmes were instigated
to obtain the redshifts for the remaining sources using the FOS and
FOS2 instruments on the INT and WHT facilities. Using automatic,
optimal extraction techniques and line fitting procedures we obtained
an average redshift accuracy of $\sim 190\kms$ \cite[ and Oliver~\et
in preparation]{phd}, c.f. an accuracy of $\sim 250 \kms$ for the QDOT
survey (Lawrence \et, in preparation).

Of all 1931 FSS-$z$~I galaxies we currently have 1769 redshifts giving
an overall redshift completeness of 91.6 per cent; while for the FSS-$z$~II
project we have redshifts for 80.4 per cent of the galaxies. A mask has been
constructed which, as well as defining the survey boundaries, excludes
a number of sectors that have not been exhaustively followed up and
those sources lying close to the IRAS coverage gap. Upon application
of this mask the completeness statistics improve to 92.5 per cent and 86.8 per cent
respectively. Above 0.25Jy (where FSS-$z$~II suffers less from flux
incompleteness) we have 95 per cent of the FSS-$z$~I redshifts and 90 per cent for
FSS-$z$~II. Detailed numbers are listed in Table~\ref{comp}.  Of the
failures the most interesting will be those with either very faint or
no optical counterparts.  VLA maps have been obtained for all sources
in the FSS-$z$~I that are either blank within the IRAS error ellipse
(to the limit of the sky survey plates) or have very faint
identifications.  $R$-band CCD images have also been obtained for many
of these sources. These have yielded a number of further
identifications and redshifts have been obtained for several of
these. Work is continuing to obtain the redshifts for the remainder.

\begin{table}\centering
\caption{Redshift completeness statistics. The subsample has sources lying
within our `mask' excluded.  Figures in parenthesis indicate the
number of galaxies for which we have redshifts.} 
\label{comp}
\begin{tabular}{crrrrrrr}
& Total
&\multicolumn{2}{c}{galaxies}
&\multicolumn{2}{c}{subsample}
&\multicolumn{2}{c}{$S_{60}>0.25$Jy}\\
\\
 N    & 1410 & 1369 & (1253) & 1369 & (1253) &  924 &  (874)\\
 A    &  474 &  457 &  (413) &  383 &  (361) &  263 &  (251)\\
 E    &  107 &  105 &  (103) &  105 &  (103) &   71 &   (70)\\
 P    & 1416 & 1391 & (1133) & 1234 & (1062) &  921 &  (826)\\
 X    &  321 &  312 &  (236) &  224 &  (203) &  146 &  (134)\\
\hline
      & 3728 & 3634 & (3138) & 3315 & (2982) & 2325 & (2155)\\
\end{tabular}                                                          
\end{table}

This paper quantifies the large scale clustering seen in both surveys
and directly compares these samples with optical surveys.  These
surveys are also being used for many other studies.  It was during the
first of these projects that the unique object F10214+4724 was
discovered \cite{f10214}.  The completeness and reliability of the
FSS-$z$~I survey will be detailed in future papers (Lonsdale \et \&
McMahon \et, in preparation).  Simple tests reveal significant
evolution within this sample \cite{hx94}.  This evolution will be
elaborated in greater detail by Broadhurst \et (in preparation). The
appearance of the Bo\"{o}tes void which overlaps with areas A \& E
will be discussed by Oliver \et (in preparation). The data for both
surveys will be presented by Oliver~\et~(in preparation).

\section{Counts-in-Cells}

To assess the large scale structure seen in the FSS-$z$ surveys
quantitatively we have chosen to apply the counts-in-cells method
described by Efstathiou \et \shortcite[; hereafter E90]{e90} and also
used by Loveday~\et~\shortcite{lday_cincell}.  The two-point 
autocorrelation function is a useful descriptive statistic; indeed for
Gaussian fields it provides a complete statistical profile of the
field.  The counts-in-cells method measures a volume integral of this
function, the variance, and avoids the problem of assigning weight to
galaxies in direct measurements of the two-point correlation function
This analysis is also relatively straightforward to perform in the
presence of a mask covering a significant fraction of the sky, a
considerable advantage over direct determinations of the
power-spectrum in Fourier space.

\subsection{Basic Method}

We consider the moments of galaxy counts ($N_i$) over a lattice of $M$
cells. Following E90 we define the statistics
\begin{eqnarray}
\bar{N} & = & \frac{1}{M}\sum_i N_i\\
S       & = &  \left(\frac{1}{(M-1)}\sum_i(N_i-\bar{N})^2\right)-\bar{N}.
\end{eqnarray} 
The statistic $\bar{N}$ has an expectation value $nV$, where $n$ is
the mean galaxy density and $V$ is the cell volume. {\em If} the cells
are independent then the expectation value of $S$, $\langle S\rangle
=n^2\sigma^2V^2$, where $\sigma^2$ is a volume integral over the
two-point autocorrelation function:

\begin{eqnarray}
\sigma^2(l) & = & \frac{1}{V^2}\int_{V_1V_2=l^3}\xi(r_{12})dV_1dV_2.
\end{eqnarray}
$\sigma^2$ is also closely related to the power spectrum as described in
Section~\ref{pspec}.

If the cells are independent and the number of cells is large then we can
derive an expression for the variance of $S$ [$\var(S)$] in terms of the
first four moments of the density field (equation~4 in E90).  
However, if we assume Gaussian fluctuations then the 
higher order reduced correlation functions vanish giving  
\begin{eqnarray}                                          
\var(S)=\frac{2n^2V^2(1+\sigma^2)+4n^3V^3\sigma^2+2n^4V^4\sigma^4}{M}.
\label{var2}
\end{eqnarray}
This allows us to estimate $\var(S)$ for a given sample by using the
empirical estimates $nV\approx\bar{N}$ and $\sigma^2\approx
S/\bar{N}^2$ in equation~\ref{var2}.  For non Gaussian fluctuations
this underestimates the errors and ideally we should use N-body
simulations of a particular model to determine higher order moments
and thus estimate the errors.

A survey volume is divided up into concentric shells and these are then
divided into $M_i$ cells.  The mean number density of the cells is
constant within a given shell allowing density estimations of the
variance from each shell.  These estimates can be combined using the
maximum likelihood technique. The statistics $N_i$ and $S_i$ are
calculated for each shell and we construct the likelihood function
from all shells

\begin{eqnarray}
{\cal L}(\sigma^2)=\prod_{i}\frac{1}{[2 \pi \var(S_i)]^{1/2}}
\exp \left[-\frac{(S_i-n_{i}^{2}V^2\sigma^2)^2} {2\var(S_i)}\right].
\label{like}
\end{eqnarray}
We can now estimate $\sigma^2$ by replacing $n_i^2V^2$ with
$\bar{N}_i^2$ and maximizing the likelihood function numerically. We
can justify not using a joint likelihood function for the two
estimators, $\bar{N}_i^2$ and $S_i$, since the errors in $\bar{N}_i^2$
are much smaller.  The 68  per cent confidence limits to this maximum
likelihood solution are the values of $\sigma^2$ where the log
likelihood has dropped by a factor of 0.5 from its maximum value.
This maximum likelihood technique automatically weights the shells,
shells with few cells or sparse density receiving less weight.

To determine how well the data are modelled by our maximum likelihood
solution we also calculate the $\chi^2$ of each fit,

\begin{eqnarray}
\chi^2=\sum \frac{(S_i-\bar{N}_i^2\sigma^{2})^2}{\var(S_i)}
\label{chisq}
\end{eqnarray}
where the number of degrees of freedom ($\nu$) is one less than the
number of shells.

\subsection{Detailed Method}

In this Section we discuss finer points of the counts in cells method.
Specifically we propose two refinements to the method of E90.

In practice each cell does not have exactly the same volume owing to the
non-uniform mask.  We correct the estimators $\bar{N}$ and $S$ for
incomplete volumes as prescribed by equation~9 in E90 (N.B. equation~9
in E90 has a minor typographical error) i.e.
\begin{eqnarray}
\bar{N} & = & \frac{\sum N}{\sum V} \\
S       & = & 
\frac{ 
\sum \left(N-\bar{N}V\right)^2 - 
\left[1-\frac{\sum V^2}{\left(\sum V\right)^2}\right]\sum N
}
{\bar{N}^2\left[
\sum V^2 - 2\frac{\sum V^3}{\sum V} + 
              \frac{ \left(\sum V^2\right)^2 }{ \left(\sum V\right)^2 }
\right]
}.
\end{eqnarray}
This correction only takes into account the fact that the expected
number of galaxies in a restricted cell will be reduced.  In addition,
a masked cell will, in general, sample smaller scales.  Since it is
observed that clustering is stronger on smaller scales this means that
the expected $\sigma^2$ increases.  This bias is difficult to correct
for since it depends in detail on the topology of the mask and the
slope of the correlation function. We have elected simply to exclude
those cells with more than half the volume masked.

At large distances the number of galaxies per cell has dropped
significantly so that we no longer get a meaningful addition to our
estimate of the variance.  E90 showed that if $\bar{N}\sigma^2\gg 1$
the variance is inversely proportional to the volume. Similarly it can
be shown that if $\bar{N}\sigma^2\ll 1$ and $\bar{N}\ll 1$ the
variance on $\sigma^2$ actually increases with additional cells.  Thus
$\bar{N}\sigma^2=1$ represents some measure of useful depth of a
survey (not surprisingly this is where the Poisson variance equals the
clustering variance).  Although the maximum likelihood analysis
includes a weighting scheme there is little point in taking shells
much beyond this limit. We estimate this depth using $\sigma^2$ from
E90 and the luminosity function of Saunders~\et~\shortcite{s90} and
only consider shells to roughly twice this depth.

It has been claimed that the counts-in-cells method as it stands is
independent of the galaxy selection function [E90, Efstathiou
\shortcite{e95}].  This is not strictly true.  Since the selection
function is not constant across each shell the galaxies will tend to
be concentrated towards the nearest side.  This reduces the effective
volume of the cells and (as with masking) will bias the variance
upwards. With prior knowledge of the clustering strength and selection
function one could estimate and remove this bias.  Instead of this we
eliminate this bias by volume limiting each shell (i.e. removing all
galaxies that would not have been visible if placed at the far side of
their shell).  The removal of galaxies should not seriously affect our
final errors since they are not sensitive to the mean density.  As
expected this correction reduces the variance in almost every case.  For
QDOT in particular the reduction was considerable, up to 1.7 times the
quoted errors.  Comfortingly the $\chi^2$ usually decreased, showing
that the fit was better.

We found that the variances obtained by the counts-in-cells method
were quite sensitive to the specific grid pattern used.  To reduce
this sensitivity we use 16 different grids for each survey.
First we divide the volumes into equal thickness concentric shells,
then divide these shells with latitude and longitude cuts such that
the resulting cells have equal volumes (in the absence of any mask).
These latitude and longitude cuts are then offset by 1/2 and 1/4
spacing in both dimensions to produce eight different grids.  This
entire process is then repeated with the shells offset by 1/2 a shell
thickness.  Each grid gives us 1$\sigma$ upper and lower limits to the
variance.  We take the linear average of these over the 16 grids
to produce a final estimate of the upper and lower limits on the
variance.  In the process we also produce an estimate of the scatter
between the individual grids; in many cases it is comparable to our
quoted errors.  By averaging over 16 different grids we have thus
undoubtedly reduced the errors.  It would be possible to account for
the reduction in systematic errors by subtracting the grid error
(estimated from the scatter between different grids) in quadrature
from the maximum likelihood error estimate.  However, in some cases
our grid error is actually larger than our maximum likelihood error,
and this procedure would lead to negative variances.  This is a
further indication that the maximum likelihood errors are
underestimated.  So a correction for the reduction in systematic
errors due to the re-gridding is not possible without using full
non-Gaussian error estimates.  Figure~\ref{grid} shows $\sigma^2$ for
each of the 16 grids used for our FSS-$z$ surveys together with
the final averaged results.

To assess the goodness of the maximum likelihood solutions we also sum
up the $\chi^2$ and $\nu$ over all grids. Since variances determined
from the shells of one grid are {\em not} independent from the shells
in another grid we have overestimated the number of degrees of
freedom, hence the overall goodness of fit will be optimistic.

In equation~\ref{like} we estimate $n_i^2V^2$ with the biased
estimator $\bar{N}_i^2$.  We have investigated replacing $\bar{N}_i^2$
with the unbiased estimator $[(\sum_i N)^2 - \sum_i N^2]/M(M-1)$ but
did not find any significant differences.

A further bias arises because the cells are not independent.  As
discussed by E90 and Loveday~\et~\shortcite{lday_cincell} this affects
the estimate of $\var(S)$.  However, correlations between cells also
bias the estimator $S$ downwards.  Ideally each shell should include
many uncorrelated volumes.  A small area survey will have fewer
uncorrelated cells per shell than a large area survey of the same
volume.  We thus might expect our FSS-$z$ to show less variance than
the all sky surveys and  since we do not see such an effect we assume that
this bias is small.

\begin{figure}
\vbox to2.5in{\rule{0pt}{2.5in}}

\caption
{$\sigma^{2}_{i}$ as a function of cell dimensions ($l$) for each of
the 16 grids of the FSS-$z$ surveys (error bars not plotted).
Also shown is the average of these 16 with averaged error bars}
\label{grid} 
\end{figure}

\subsection{Results}

We apply the above method to our new surveys and also to existing IRAS
redshift surveys: the latest version of the QDOT survey
(Lawrence~\et~in preparation); the 1.2~Jy survey \cite{fish_data}; a
faint QDOT sample (0.6-1.2~Jy); a combined PSC (QDOT and 1.2~Jy)
sample. We also combine our FSS-$z$ sample with the combined PSC sample to
produce a single IRAS sample.

For our FSS-$z$ surveys we flux limit the whole sample at 0.25Jy to
allow for the varying completeness limits in different areas.  The
mask was described in Section~\ref{intro}.

For the QDOT and faint QDOT samples we use the standard QDOT mask.
The faint QDOT sample has $0.6<S_{60}<1.2$ Jy so the volume limiting
is rather harsh.

In the analysis of the 1.2~Jy sample we apply a cut of $|b|>5\deg$
together with the QDOT mask. This is slightly more severe than the
1.2~Jy team used themselves.

Since the faint QDOT sample has no galaxies in common with the 1.2~Jy
survey the two can be combined to construct a larger PSC sample.  For
this PSC sample we use the same mask as we used for the 1.2~Jy
catalogue.  Many fewer galaxies are excluded when volume limiting the
combined PSC sample than are excluded when volume limiting the samples
separately.

Finally, we take the individual variance estimates from each shell of
our combined PSC sample, together with the individual variances from
each shell of our FSS-$z$ sample, to construct a single likelihood
function (equation~\ref{like}), as before.  Maximizing this likelihood
function gives us a single variance estimate from all samples.  Since
the volume of overlap between the FSS-$z$ and the PSC is small, any
interdependence will not be a significant problem.

The results for all the various samples described above are given in
Table~\ref{cincell_results}.

In many cases the $\chi^2$ and $\nu$ values indicate that the goodness
of fit for the individual grids is poor.  As discussed in E90 the
assumption of Gaussian fluctuations causes the errors to be
underestimated which may explain this.  Since we reduce the systematic
errors by averaging over 16 grids, without adjusting our error
estimates accordingly we have to some extent compensated for this.

\begin{table}\centering
\caption{$\sigma^2$ for the various samples, 
averaged over 16 grids in each case.  Including $\chi^2$ and grid
scatter estimate}
\label{cincell_results} 
\begin{tabular}{ccrrrr}

$l$ & $\sigma^2$ 
& \multicolumn{1}{c}{$\chi^2$} 
& \multicolumn{1}{c}{$\nu$} 
& \multicolumn{1}{c}{$P(\chi^2,\nu)$} 
& \multicolumn{1}{c}{$\sigma_{\rm GRID}$} \\
 \\ \multicolumn{6}{c}{FSS-$z$ Samples} \\ \\ 
 10& $    0.863\pm    0.097 $ &   296 &   251&     2.762&     0.056 \\
 20& $    0.322\pm    0.059 $ &   338 &   251&     0.020&     0.050 \\
 30& $    0.172\pm    0.053 $ &   402 &   235&     0.000&     0.070 \\
 40& $    0.182\pm    0.070 $ &   320 &   203&     0.000&     0.063 \\
 60& $    0.095\pm    0.053 $ &   105 &   107&    53.387&     0.087 \\
 \\ \multicolumn{6}{c}{QDOT Sample} \\ \\ 
 10& $    0.716\pm    0.108 $ &   133 &   104&     3.026&     0.073 \\
 20& $    0.308\pm    0.061 $ &   123 &    88&     0.831&     0.074 \\
 30& $    0.199\pm    0.054 $ &   137 &    72&     0.001&     0.038 \\
 40& $    0.127\pm    0.041 $ &   111 &    72&     0.239&     0.048 \\
 60& $    0.069\pm    0.030 $ &    34 &    40&    73.224&     0.045 \\
 \\ \multicolumn{6}{c}{QDOT Faint($S_{60}<1.2$ Jy) Sample} \\ \\ 
 10& $    0.811\pm    0.365 $ &   111 &    72&     0.209&     0.432 \\
 20& $    0.709\pm    0.288 $ &    72 &    56&     7.358&     0.286 \\
 30& $    0.392\pm    0.311 $ &    46 &    40&    24.406&     0.391 \\
 40& $    0.152\pm    0.337 $ &    28 &    40&    92.946&     0.272 \\
 \\ \multicolumn{6}{c}{1.2 Jy Sample} \\ \\ 
 10& $    0.864\pm    0.067 $ &   225 &   168&     0.216&     0.058 \\
 20& $    0.310\pm    0.039 $ &    96 &   120&    94.926&     0.031 \\
 30& $    0.156\pm    0.031 $ &   120 &   104&    13.782&     0.017 \\
 40& $    0.095\pm    0.025 $ &   150 &    88&     0.004&     0.019 \\
 60& $    0.057\pm    0.022 $ &    33 &    40&    78.327&     0.033 \\
 \\ \multicolumn{6}{c}{PSC Samples} \\ \\ 
 10& $    0.889\pm    0.059 $ &   179 &   184&    58.199&     0.055 \\
 20& $    0.322\pm    0.035 $ &   124 &   136&    75.696&     0.023 \\
 30& $    0.152\pm    0.027 $ &    96 &   104&    71.215&     0.021 \\
 40& $    0.094\pm    0.023 $ &   123 &    88&     0.817&     0.014 \\
 60& $    0.048\pm    0.018 $ &    36 &    24&     5.867&     0.028 \\
 \\ \multicolumn{6}{c}{PSC + FSS  Samples} \\ \\ 
 10& $    0.881\pm    0.050 $ &   479 &   451&    17.886&     0.041 \\
 20& $    0.322\pm    0.030 $ &   469 &   403&     1.281&     0.023 \\
 30& $    0.154\pm    0.024 $ &   519 &   355&     0.000&     0.022 \\
 40& $    0.107\pm    0.024 $ &   517 &   307&     0.000&     0.012 \\
 60& $    0.055\pm    0.019 $ &   205 &   147&     0.109&     0.029 \\

\end{tabular}  
\end{table}

The QDOT variances appear slightly smaller than those quoted by E90 and
Efstathiou \shortcite{e95}.  The main reason for this is that by
volume limiting each shell we have removed the bias due to gradients
in the selection function.  The grids also play a part in this
apparent discrepancy; without volume limiting, one grid did produce a
variance as high as the E90 variance at $40h^{-1}$ Mpc. Notice that
the scatter between grids is always comparable to our quoted error
(i.e. roughly the maximum likelihood error from a single grid).  The
explanation for the high E90 variance at $40h^{-1}$ Mpc given by
Efstathiou \shortcite{e95} is that there was an upward statistical
fluctuation in the number of QDOT galaxies in Hercules.  We believe that
this fluctuation had a bigger impact on the E90 analysis than on ours
partially because we used multiple grids and partially because we
removed the selection function bias.

The 1.2~Jy results, on the other hand, appear slightly larger than the
results of Fisher \et \shortcite{fish_lss}. This may be due to
differences in the masks.

In comparing estimates of clustering from different surveys we can
assume that the ratio of variances is independent of scale and use a
weighted average of the log of this ratio (with associated $\chi^2$
and $\nu$) to estimate $b_a/b_b=\sqrt{\sigma^2_{\rm a}/\sigma^2_{\rm
b}}$.  We find $b_{\rm QDOT}/b_{\rm 1.2~Jy}=0.98\pm0.06$, $b_{\rm
FSS}/b_{\rm 1.2~Jy}=1.03\pm0.06$ and $b_{\rm FSS}/b_{\rm QDOT}=1.06\pm0.07$.  
All have acceptable $\chi^2$ values (indicating
that the scale independence is justifiable) and on each scale the
variances between surveys are consistent.  All are consistent with
$b_a/b_b=1$ and in addition there is no significant disagreement
between the variances measured by the different surveys on any scale.
This can be seen in Figure~\ref{fig_sig}.  This in turn justifies our
combining the samples.  (In these comparisons we have assumed that the
errors from each survey are independent. The many galaxies in common
between QDOT and 1.2~Jy suggest that they are not in fact independent so we
have overestimated the errors in $b_{\rm QDOT}/b_{\rm 1.2~Jy}$.
However, the true errors would need to be much smaller for the
the surveys to disagree significantly.)

The agreement between QDOT and 1.2~Jy is slightly at odds with the conclusions
of Tadros \& Efstathiou's \shortcite{te95} power spectrum analysis of
the two surveys.  They find a discrepancy between the two surveys
which disappears after the exclusion of the Hercules region.  Their
analysis uses a selection function derived from the 60$\mu$m
luminosity function \cite{s90}.  It is possible that their analysis
is more sensitive to the Hercules anomaly as a result of errors in this
selection function.  At the distance of Hercules, errors in the
evolutionary term of the Saunders~\et \shortcite{s90} luminosity
function contribute a systematic 7.5 per cent uncertainty to their selection
function.  Our analysis is independent of the selection function.
Differences between the mask used by Tadros \& Efstathiou 
and the one we use could also have affected this comparison,
by excluding significant features in one or other analysis.

A more detailed comparison of the 1.2~Jy survey and QDOT survey
shows that they are consistent with a single underlying density
field \cite{e95}.

We can apply the same test to compare the variances of these surveys
with the standard CDM variances quoted by E90.  For none of the
surveys does the weighted ratio of galaxy to CDM variances provide
an acceptable fit at around 99 per cent significance.  Using the full IRAS
sample we find a weighted average $b_{\rm IRAS}/b_{\rm
CDM}=0.95\pm0.02$ (this average is dominated by the smallest
scales). The fit is bad, $\chi^2=17.0, \nu=4$, allowing us to reject
the CDM model at the $>99$ per cent level. This is clearly shown in
Figure~\ref{fig_sig} where the slope of the variances with scale from
the combined IRAS data is clearly at odds with the CDM slope.

\begin{figure}
\vbox to2.5in{\rule{0pt}{2.5in}}
\vbox to2.5in{\rule{0pt}{2.5in}}

\caption{Counts in cells variance as a function of scale:
Top: 1.2~Jy Survey - solid line and stars; QDOT Survey - dashed line
and circles; FSS-$z$ - dot/dash line and crosses.  Bottom: Standard
Cold Dark Matter (E90) - sold line and stars; all IRAS surveys
together - dashed line and circles.}
\label{fig_sig} 

\end{figure}

\subsection{Comparison With Model Power Spectra}
\label{pspec}

Theoretical clustering predictions in the linear regime are frequently
expressed in terms of the power spectrum $P(k)$ or in dimensionless
form $\Delta^{2} \propto k^{3}P(k)$.  The quantity $\Delta^{2}$ is
defined as the variance per $\ln k$ i.e.  $\Delta^2 \equiv
d\sigma^2/d\ln k$.  For a power law spectrum ($\Delta^{2} \propto
k^{n+3}$) the fluctuations within Gaussian spheres (radius $R_G$) is
given by
\begin{eqnarray}
\sigma^{2} & = & \Delta^{2}(k)\\
k          & = &\left[\frac{1}{2}(\frac{n+1}{2})!\right]^{1/(n+3)}\frac{1}{R_G}.\label{pk1}
\end{eqnarray}
For cubical cells ($V=l^3$) the same expression applies but with
$R_G\rightarrow l/\sqrt{12}$ \cite{jp91}.  The theoretical predictions
relate to the real-space mass fluctuations whereas the $\sigma^2$ we
have calculated is for galaxies in redshift-space.  A common
assumption is that the real-space galaxy and mass variances are
linearly related by a bias parameter $b$ (see
e.g. equation~\ref{bias_equ}, below)
i.e. $\sigma^{2}_{m}=\sigma^{2}_{g}/b^2$, while the 
redshift-space/real-space distortions on large scales can be corrected for using the
relation
\begin{eqnarray}
\sigma^{2}_{\rm real}=\sigma^{2}_{z}
\left[1+\frac{2}{3}\left(\frac{\Omega^{0.6}}{b}\right)
+\frac{1}{5}\left(\frac{\Omega^{0.6}}{b}\right)^2\right]
\label{pk2}
\end{eqnarray}
\cite{nk_zspace}.

We also make a functional correction to the model power spectrum to
provide an approximation to the non-linear evolution of the power
spectrum \cite{jp_dodds}.  In addition we use the Peacock \& Dodds
\shortcite{jp_dodds} correction for the redshift-space distortions
which accounts for small scale `finger of god' effects (for this we
assumed a pairwise velocity dispersion of 450 kms$^{-1}$ added in
quadrature with the redshift measurement error of 190 kms$^{-1}$).
	
Figure \ref{p_k} shows the predicted Fourier amplitudes from the FSS-z
sample, QDOT, the 1.2~Jy sample and the PSC-FSS sample, calculated
from equations \ref{pk1} and \ref{pk2} and using the results
from Table \ref{cincell_results}. We also show the expected results
from the APM power spectrum inversion of Baugh \& Efstathiou
\shortcite{be94}. We have assumed that the linear IRAS bias is $b_I=0.7$
and the optical APM galaxies have a bias of $b_o=0.9$, consistent
with the results of Section 3, below.

These results are compared with the Mixed Dark Matter (MDM) model,
which has previously been shown to be in good agreement with large
scale clustering data \cite[Schaefer \& Shafi 1992]{ant_rr}.  The
model shown has $\Omega_{HDM}=0.15$ and $h=0.5$, taken from the results
of van Dalen \& Shaefer \shortcite{mdm}, and is normalized to the COBE
spectrum assuming $n=1$, yielding an effective quadrupole of $19.9 \pm
1.6 \mu$K derived by Gorski \et \shortcite{gorski}. We have also
plotted the scale invariant $n=1$ Harrison--Zeldovich spectrum.

Figure \ref{p_k} again shows that there is now broad agreement between
QDOT, 1.2~Jy and the PSC-FSS surveys.  The MDM model is a good fit to
the APM data, although it appears to have a slightly steeper slope
than the IRAS data.  This fit could not be improved by increasing the
proportion of hot dark matter (HDM) because, with the same COBE normalization, we would
then predict too much power on all these scales.  Improvements could
be made, either by adjusting the primordial spectral index, or by
introducing a scale dependent bias.  There are a number of reasons why
the small discrepancies between the MDM model and the IRAS data do not
justify such fine tuning.  First, there are a number of debatable
assumptions that have been made in order to predict the non-linear,
redshift-space power spectrum from a linear theory, real-space power
spectrum.  Secondly, the assumption of Gaussian fluctuations in the data
analysis causes us to underestimate our errors as described above.
Thus we would conclude that our data are not incompatible with the MDM
model.

\begin{figure}
\vbox to2.5in{\rule{0pt}{2.5in}}
%

\caption{
Square root of the Power spectrum, $\Delta(k)$, for the FSS-$z$
(stars), QDOT (circles). 1.2~Jy (triangles) PSC-FSS (squares) and APM
(open circles).  The data points have been offset along the $k$
axis for clarity. The curved solid line is the Mixed Dark Matter model
with 15  per cent hot dark matter and $h=0.5$, normalized to COBE, and
corrected for nonlinear effects.  The broken line is the linear
spectrum.  The straight solid line is the extrapolated scale invariant
Harrison-Zeldovich spectrum. The data has been corrected for linear
redshift space distortion in the small angle regime, and a linear bias
of $b_I=0.7$ for the IRAS galaxies and $b_O=0.9$ for the optical APM
galaxies.
}
\label{p_k} 
\end{figure}

\section{Comparison With Optical}
\label{comp_intro}

An interesting question in the study of large scale structure is how
the clustering properties of different galaxy populations are related.
Since the various structure formation theories such as CDM predict the
mass distribution, whereas redshift surveys  probe only the galaxy
distribution, we need to know how these are related.  If different
classes of galaxies do not trace the same structures then clearly at
best only some of these classes can be faithfully tracing the mass.
With an increased understanding of how different classes are related
we can hope to gain some understanding of how the galaxies relate to
the mass. This question is particularly pertinent to the
interpretation of flux limited redshift surveys where the mix of
populations changes with both the flux limit and the redshift.  Any survey
that contains two (or more) populations with different clustering
properties is inevitably going to produce ambiguous results dependent
on selection functions of the classes involved and the scales being
considered.

A very simple possible relationship between galaxy populations is that
they linearly trace some underlying density field subject to finite
sample errors, i.e.
\begin{eqnarray}
\frac{1}{b_a}\frac{\delta \rho_a}{\bar{\rho_a}}({\bf r})=
\frac{1}{b_b}\frac{\delta \rho_b}{\bar{\rho_b}}({\bf r})=
\frac{1}{b_c}\frac{\delta \rho_c}{\bar{\rho_c}}({\bf r}), {\rm etc}.
\label{bias_equ}
\end{eqnarray}
(If the galaxy distributions were identical then obviously $b_a\equiv
b_b$ etc..) A more general relationship could be expressed as
\begin{eqnarray}
\frac{\delta \rho_a}{\bar{\rho_a}}({\bf r})=
f\left(\frac{\delta \rho_b}{\bar{\rho_b}}({\bf r})\right)
\end{eqnarray}
in which case the linear-biasing model would be the first order
approximation in the case of small $\frac{\delta \rho}{\bar{\rho}}$.
Under the assumption of linear-biasing one can use the ratio of
clustering amplitudes and similar statistics for different galaxy
classes to estimate the relative bias parameters [e.g.
$\sigma^{2}_a/\sigma^{2}_{b}=(b_a/b_b)^2$].

Comparisons of optical and infrared galaxy clustering have been made,
but give conflicting answers.  Babul \& Postman \shortcite{bpost90}
compare IRAS galaxies and CfA galaxies in the first CfA slice and
conclude that they are consistent with same underlying galaxy
distribution outside the core of the Coma cluster (i.e. $b_O/b_I=1$).
The ratio of variances between the QDOT and APM-Stromlo surveys,
$\sigma^2_O/\sigma^2_I=1.0\pm0.35$ \cite{lday_cincell} implies
$b_O/b_I=1.0\pm0.18$ (95 per cent confidence), while power spectra analysis
suggest $b_O/b_I=1.3$ \cite{jp_dodds}.  Dynamical estimates of
$\beta\equiv \Omega^{0.6}/b$ from IRAS samples lie between
$\beta_I=0.6$ and $\beta_I=1.3$ with excursions to $\beta_I=0.25$;
optical estimates lie between $\beta_O=0.4$ and $\beta_O=0.75$ with
earlier estimates as low as $\beta_O=0.1$ \cite{dek94}.  The ratio
$\beta_I/\beta_O=b_O/b_I$ but the wide range of estimates allows $0.8<
b_O/b_I<3.3$.  Ratios of cross- and auto-angular-correlations between
the UGC ESO and IRAS suggest $b_O/b_I=1.4\pm0.1, b_O/b_I=1.0\pm0.3$ or
$b_O/b_I=2.0\pm0.6$ depending on which correlation functions are
compared \cite{ol_angb}.  On small scales the 2 Jy/CfA
cross-correlation amplitudes suggested $b_O/b_I=2$, while $b_O/b_I=1$
was found when comparing the 2Jy survey with the Southern Sky Redshift
Survey \cite{2jy_lss}.  Fisher~\et \shortcite{fish_lss} found
$b_O/b_I=1.38\pm0.12$ when comparing $\sigma_8$ from the 1.2~Jy survey
with the optical $\xi(r)$ of Davis \& Peebles \shortcite{d_peb}.

We can use our estimates of the variance in cells together with that
from an optical survey to estimate $b_O/b_I$.  Using our PSC + FSS-$z$
sample with the APM-Stromlo variances in cubic cells gives
$b_O/b_I=1.20\pm0.05$ with $\chi^2=1.2, \nu=4$.  This is consistent
with some of the lower determinations above but rules out $b_O/b_I=1$.

There are two problems with these analyses.  First they do not test
the underlying hypothesis that the distributions are linearly biased
with respect to each other. Secondly the IRAS and optical classes are
not clearly defined galaxy classes. What constitutes an IRAS
galaxy depends on both the flux limit used and the redshift at which
you find it, and the same is true of optical galaxies.  Most notably a
given galaxy can be seen in both optical and IRAS surveys.

In this section we are going to explore the relationship between the
clustering of IRAS and optical galaxies taken from our FSS-$z$ surveys
and the CfA samples.  By design, the FSS-$z$ surveys overlap with the
first three strips of the CfA 2 survey.  We are thus in a position to
compare density fields directly point-by-point, rather than by
comparing global clustering statistics.  We will define IRAS and
optical galaxies such that the classes are unique (no optical galaxy
can also be an IRAS galaxy) and on the basis of a physical distinction
that (to first order) is independent of redshift.

\subsection{The CfA and FSS-$z$ Samples}
\label{comp_sample}

The optical parent catalogue is the only published strip in the
second CfA survey \cite{cfa}; i.e.  $26.5\deg < \delta < 32.5 \deg,
8^{\rm hr} < \alpha < 16 ^{\rm hr}$.  The IRAS  parent catalogue
comprises the FSS-$z$ samples discussed above, including all galaxies down to
0.2 Jy.

Since our aim is to compare the structures traced by different
galaxies within the same volumes we must first restrict the two parent
surveys to have the same areal profile.  This means excluding CfA
galaxies which lie within our FSS-$z$ mask and also any FSS-$z$
galaxies outside the CfA boundaries.  These cuts leave us with 976
FSS-$z$ galaxies and 854 CfA galaxies.

Of course many galaxies that appear in the CfA catalogue will also be
found in the FSS-$z$.  We identify all such pairs using a 3\arcmin \
search radius.  If more than one partner is found for a given source
the match is made on the basis of brightness, nearness etc. (the
details of the duplicate removals are relatively unimportant since most
duplicates are neighbouring galaxies).  One of these pairs was found
to have a much larger redshift in one catalogue and, since the angular
separation was also large, this match was rejected as a chance
projection.  This matching left us with three catalogues: 554 galaxies
seen {\em only} in the CfA 2 survey; 300 galaxies seen in both
surveys; 676 galaxies seen {\em only} in the FSS-$z$ surveys.  These
three catalogues are statistically independent, in the sense that no
galaxy appears in more than one sample.

We now require a more physically meaningful classification of `IRAS'
and `optical' galaxies.  The mix of galaxy types in our three samples
depends to a large extent on the survey selection functions, for
example a nearby elliptical may be seen in both surveys but a similar
galaxy would be unlikely to be detected in the FSS-$z$ sample at a
greater distance.  To obviate this problem we have introduced a $60
\mu$m, $B$ magnitude colour cut which neatly divides the common sample
into two.  The colour chosen is simply the ratio of the respective
survey limits i.e.
\begin{eqnarray}
\frac{S_{60}}{0.2}=10^{0.4(15.5-m_B)}.
\end{eqnarray}
Naturally all objects seen only in the FSS-$z$ survey will have a
colour that is more infrared than this limit and vice versa.  The only
dependence on redshift that is now present in our classification
enters through the difference between the slopes of the spectral
energy distributions around the $60\mu$m and $B$-bands.  Applying
reasonable K-corrections at $z=0.3$ this flux criterion corresponds to
a cut at $L_B>2L_{60}$ and although this cut is essentially arbitrary,
it provides a much more appealing definition of an IRAS galaxy or
optical galaxy than its presence or absence in a flux limited
catalogue.

After applying this cut we arrive at two samples:
\begin{description}
\item 866(735) IRAS galaxies,
\item 664(639) optical galaxies,
\end{description}
where the numbers in parentheses indicate those sources that have
redshifts.

We already know that elliptical galaxies do not trace the same density
field as other galaxies \cite{dressler_morph} and we also know that
ellipticals are under-represented in IRAS sample \cite{dej84}.  So a
priori we would expect some difference between the clustering of these
two surveys simply from this morphological effect.  This will be
particularly important in this comparison since the CfA strip is
strongly dominated by the Coma cluster.  We therefore create a further
sample in which we restrict the optical galaxies to exclude all
elliptical galaxies.  To identify the ellipticals we use the
morphological classifications in the CfA catalogue and exclude all
types earlier than Sa.  This gives us an optical-spiral sample with
269(267) galaxies.

\subsection{Qualitative comparison}

The Figure~\ref{cone} shows the cone diagrams for the three subsamples
described in Section~\ref{comp_sample}.  In the optical sample the
most striking structure is the Coma cluster and `Great Wall'.  In the
IRAS sample the Coma cluster is considerably less pronounced, although
the `wall' itself appears equally prominent.  The difference in
appearance of the Coma cluster between the two samples is easily
explained; the optical sample contains more ellipticals which are
preferentially found in rich clusters. Clearly the IRAS sample probes
much deeper than the optical sample although they have similar numbers
of galaxies (this is because of the breadth of the IRAS luminosity
function) i.e. the IRAS galaxies are sparser.  There is some
indication of a second `wall' at around 15000\kms in the IRAS sample.
Although this feature is poorly sampled in the optical at this depth,
nevertheless with the IRAS galaxies to guide the eye one can pick it
out.  Beyond 10000 \kms between $12^h$ and $13^h$ there  appear to
be some voids of low signal-to-noise ratio in the optical data,
 but these are much
less pronounced or non-existent in the IRAS data.  In the
optical-spiral sample we see that the Coma cluster is far less
significant (as we would expect).  More detailed observations are
strongly limited by the sparsity of the galaxies.

\begin{figure}                                       
\vbox to7.5in{\rule{0pt}{7.5in}}
\caption{Cone Diagram of FSS-$z$/CfA2 Strip 1 intersection.
Top -- IRAS Galaxies.  Middle -- optical galaxies.
Bottom -- optical-spiral galaxies.}
\label{cone}
\end{figure}

Compressing the cone diagrams in the tangential dimension increases
the signal-to-noise ratio at larger radii (Figure~\ref{nz}). Here we can
more clearly see the second wall in the IRAS sample and guided by that
we see a second peak in the optical data.  Notice that the peak
corresponding to the `Great Wall' is a single broad peak in the
optical but has two peaks in the IRAS data. This arise from
the difference in selection functions.

\begin{figure}
\vbox to7.5in{\rule{0pt}{7.5in}}
\caption
{$N(z)$ diagrams.  The top diagram shows the $N(z)=\rho(z)dV$ for the
IRAS sample while the middle diagram shows the same for the optical
Sample, the bottom diagram is the ratio between the two (together with
a power-law fit) this gives us $\bar{\rho_I}/\bar{\rho_O}$ as a
function of $z$, independent of clustering (if there is no relative
biasing)}
\label{nz} 
\end{figure}

The cone diagrams above are dominated by the selection functions of
the various samples.  Dividing the volume into cells within concentric
shells we can estimate $\rho/\bar{\rho}=n_i/\sum n_i$ for each class
within each cell.  In Figures~\ref{density1},~\ref{density2} 
we have plotted the
densities of each cell as seen in the optical sample against the
densities as seen in the IRAS sample.  Since there is no overlap between
cells all the data points are independent.  We have not plotted error
bars which would have confused the plot.  It should be noted that the
under-densities will usually have fewer galaxies and therefore larger
errors.  The general trend in these plots seems to be a steeper slope
than would be expected for identical clustering, or even a linear-bias
model with $b_O/b_I=1.3$.  On larger scales this appears to be true
both above and below the mean density indicating that, in the optical,
not only are the clusters more dense but also the voids are less dense.  On
scales of $5h^{-1}$ Mpc the IRAS density appears to be higher than the
optical density for densities larger than the mean density. This might
be explained by fingers of god which would tend to smear out the
optical clusters.

\begin{figure}
\vbox to7.0in{\rule{0pt}{14.0in}}
\caption{$\rho_O/\bar{\rho_O}\; {\rm v}\; \rho_I/\bar{\rho_I}$ diagram of
FSS-$z$/CfA2 Strip 1 intersection.  Densities are calculated in nearly
cubical cells of sizes 5 (top) and 10 $h^{-1}$ Mpc (bottom).  The
straight line would indicate identical clustering, the curved line
would be followed in a linear-biasing model with $b_O/b_I=1.3$.  All
points are independent so the scatter is real, to avoid confusion we
have not plotted error bars.}
\label{density1}
\end{figure}

\begin{figure}
\vbox to7.0in{\rule{0pt}{14.0in}}
\caption{As for Figure \protect\ref{density1} but for scales
20 (top) and 30 $h^{-1}$Mpc (bottom).}
\label{density2}
\end{figure}

\subsection{Contingency Table}

We now turn to a quantitative comparison between the two samples.
This method was first discussed by Oliver~\shortcite{phd}.
Recently a similar but statistically more rigorous method
was brought to our attention \cite{e95}.

\subsubsection{Method}
\label{cont_method}

The most obvious scheme to test is that the IRAS and optical galaxies
cluster in {\em exactly} the same way (i.e. $b_O/b_I=1$).  Our null
hypothesis is thus that the over-density in any given cell is
independen  of galaxy class.  We divide our survey volume
into shells and cells as we did for the counts in
cells analysis.  Under our null hypothesis we are at liberty to
combine the samples to give a single estimate of the `galaxy'
over-density in any cell.  From this estimated over-density and the
observed over-densities in each individual class we construct a
statistic, $X^2$, \cite{ken_stu} which we compute across the whole
shell;
\begin{eqnarray}
X^2=\sum_{ij} \frac{(n_{ij}-n_{i\star }n_{\star j}/n)^2}
                      { n_{i\star }n_{\star j}/n }.
\label{X}
\end{eqnarray}
Here $n$ ($=\sum_{ij}n_{ij}$) represents the total number of galaxies
from both surveys in a given shell.  The subscript $i$ refers to the
individual cells in the shell and $j$ to the galaxy class.  We have
employed the notation $n_{i\star}=\sum_j n_{ij}$ to represent the
marginal distributions.  Here we have implicitly assumed that the
observed galaxies give us an estimate of the underlying galaxy-density
field, and that the measured difference in densities between the two
populations is governed by Poisson statistics, i.e. that the
additional error terms due to clustering (and cross-correlations
between cells) cancel out.  In this case $X^2$ is asymptotically
distributed as $\chi^2$ and, when summed over all shells, has $M-N$
degrees of freedom where $M$ is the total number of cells and $N$ is
the number of shells.

\subsubsection{Selection Function}

As with the counts-in-cells analysis this analysis is affected by the
selection functions of the samples.  If the relative density of the
two samples $\bar{\rho_I}/\bar{\rho_O}$ changes significantly across a
shell then this will introduce spurious discrepancies between the
optical and IRAS clustering.  As before, we could volume limit each
sample within each shell.  This discards more galaxies than necessary
and for this analysis our statistics are sensitive to the mean
density.  To avoid any bias we require $\bar{\rho_I}/\bar{\rho_O}$ to
be constant over each shell.  We can estimate this quantity neatly,
without calculating either selection function independently, and (in
the case of no relative bias) independently of clustering.  It is
simply the ratio of the two $\rho(z)$ distributions as shown in
Figure~\ref{nz}.  We have fitted the observed ratio with a power law
giving us $\bar{\rho_I}/\bar{\rho_O}(z)\propto(1+z)^{27}$.  In any
given shell we then randomly throw away galaxies from the most dense
sample to give us a constant $\bar{\rho_I}/\bar{\rho_O}$ over that
shell.  This is very efficient since the errors are dominated by the
least dense sample and in any case the gradients across a shell are
small.  Our results were not significantly affected by this
correction.

\subsubsection{Cells with few galaxies}

Technically we can calculate $X^2$ for all cells where there is at
least one galaxy in each class but the $X^2$ statistic is only
distributed as $\chi^2$ in the limit of large numbers so it is unwise
to take those results at face value.  To avoid the problem of poorly
sampled cells requires some care.  If we simply exclude cells with
fewer than $n_{\rm lim}$ galaxies in any class we would end up biasing
our results.  This is because we would be excluding cells with a
higher over density in the class which has the lower mean density, and
thus concluding that the other population had more under-dense regions.
It is thus clear that any thresholding has to be imposed on $\delta
\rho/ \bar{\rho}$ rather than $n$.  We thus impose the limit
\begin{eqnarray}
\frac{n_{i\star}}{n}>\frac{n_{\rm lim}}{{\rm min}(n_{\star j})},
\label{nlim}
\end{eqnarray}
where we have replaced $\delta \rho/ \bar{\rho}$ by the estimator
$n_{i\star}/n$.  Now our limit, $n_{\rm lim}$, determines the smallest
number of galaxies of the most sparse class that we would expect in
any cell.  The actual $\delta \rho/ \bar{\rho}$ limit varies from
shell to shell, allowing us to extract the maximum information from
the surveys.  These sparse cells still contain useful information,
especially about the voids.  Fortunately our null hypothesis allows us to
group cells together.  We choose to group sparse cells with their
neighbour but one, rather than excluding them totally.  We present
results for $n_{\rm lim}= 1,2,3,4, 5$.

\subsubsection{Results}

The results of our contingency table analysis are presented in
Table~\ref{cont_table}.  The probability $P(X^2,\nu)$ is that of
obtaining a value of $X^2$ higher than that observed assuming our null
hypothesis and that $X^2$ is distributed as $\chi^2$ with $\nu$
degrees of freedom.
\begin{table}\centering
\begin{tabular}{crrrc}\\ \hline
Scale        & $X^2$ & $\nu$ & $P(X^2,\nu)$ & $n_{\rm lim}$ \\
$/h^{-1}$Mpc &       &       &  /\%         &             \\ \hline
 5 &  86.6 &   49 &  0.07470 & 1\\
 5 &  70.6 &   31 &  0.00640 & 2\\
 5 &  49.1 &   18 &  0.01036 & 3\\
 5 &  25.7 &   10 &  0.41861 & 4\\
 5 &  26.3 &    8 &  0.09185 & 5\\
 & & & & \\
10 &  86.4 &   47 &  0.04046 & 1\\
10 &  68.3 &   33 &  0.02920 & 2\\
10 &  57.4 &   27 &  0.05782 & 3\\
10 &  39.8 &   19 &  0.34293 & 4\\
10 &  44.1 &   15 &  0.01071 & 5\\
 & & & & \\
20 &  41.7 &   17 &  0.07288 & 1\\
20 &  38.4 &   11 &  0.00677 & 2\\
20 &  37.4 &    9 &  0.00226 & 3\\
20 &  30.7 &    7 &  0.00702 & 4\\
20 &  28.8 &    6 &  0.00662 & 5\\
 & & & & \\
30 &  12.8 &    7 &  7.60401 & 1\\
30 &  11.5 &    6 &  7.37997 & 2\\
30 &  12.4 &    6 &  5.32677 & 3\\
30 &  12.9 &    5 &  2.43094 & 4\\
30 &   5.7 &    4 & 22.12649 & 5\\
 & & & & \\

\hline
\end{tabular}
\caption{Contingency table analysis results.  Under the null hypothesis 
that the optically selected galaxies trace the same field as the IRAS
galaxies $P(X,\nu)$ gives the probability of obtaining $X^2$ higher
than the observed value, assuming that $X^2$ is distributed as
$\chi^2$ (the statistic $X^2$ is defined in equation~\protect\ref{X}).
$n_{\rm lim}$ is the threshold used in merging cells with few galaxies
(equation~\protect\ref{nlim}).}\label{cont_table}
\end{table}
\begin{table}\centering
\begin{tabular}{crrrc}\\ \hline
Scale        & $X^2$ & $\nu$ & $P(X^2,\nu)$ & $n_{\rm lim}$ \\
$/h^{-1}$Mpc &       &       &  /\%         &             \\\hline
 5 &  54.1 &   48 & 25.35328 & 1\\
 5 &  37.3 &   26 &  7.08115 & 2\\
 5 &  16.9 &   15 & 32.49382 & 3\\
 5 &  10.3 &    8 & 24.47746 & 4\\
 5 &   8.6 &    7 & 28.36795 & 5\\
 & & & & \\
10 &  58.1 &   37 &  1.48361 & 1\\
10 &  19.8 &   22 & 59.62626 & 2\\
10 &  25.7 &   18 & 10.63065 & 3\\
10 &  25.1 &   15 &  4.90583 & 4\\
10 &  12.4 &   10 & 25.70629 & 5\\
 & & & & \\
20 &  16.7 &   11 & 11.59919 & 1\\
20 &  15.1 &    8 &  5.67010 & 2\\
20 &   8.5 &    6 & 20.18399 & 3\\
20 &   7.1 &    5 & 21.62374 & 4\\
20 &   6.2 &    5 & 28.38374 & 5\\
 & & & & \\
30 &  15.9 &    6 &  1.41893 & 1\\
30 &  14.5 &    5 &  1.27080 & 2\\
30 &   7.7 &    2 &  2.07805 & 3\\
30 &  10.5 &    2 &  0.51919 & 4\\
30 &   7.8 &    2 &  2.02121 & 5\\
 & & & & \\
\hline
\end{tabular}
\caption{Contingency table analysis results. As for Table~\protect\ref{cont_table}
except that the CfA ellipticals were excluded prior to analysis.}
\label{cont_table_e}
\end{table}
One can see from this Table that the hypothesis that the optical
galaxies and the IRAS galaxies are identically tracing the same
underlying field is ruled out on scales $5-20h^{-1}$Mpc with a high
level of significance ($\gg 99$\%).  On the $30h^{-1}$Mpc with $n_{\rm
lim}=5$ it is impossible to rule out the null hypothesis, although the
high $\chi^2$ at smaller $n_{\rm lim}$ may hint that our statistics
are too poor for this to be informative.

As stated before, we expect some discrepancy because of the
over-representation of elliptical galaxies in optical surveys.  In
Table~\ref{cont_table_e} we present the results after having excluded
the CfA ellipticals.  Here the story is markedly different.  We are
unable to exclude our null hypothesis for scales $5, 10, 20h^{-1}$
Mpc.  At 30$h^{-1}$ Mpc we can formally rule out our null hypothesis
at the $95$ per cent level. With only two degrees of freedom this result
should be treated with caution.

We thus conclude that the detectable differences in clustering between
these two surveys are mainly due to the higher proportion of
ellipticals in the optically selected samples.  It is nevertheless
surprising that significant differences persist to large
scales where we might have expected the overabundance of ellipticals
in rich clusters to be of less importance.  This could be because of the
Coma `finger of god' which is quite extensive in redshift space.
Alternatively it might suggest that the morphology-density relation
extends to large scales.

\subsection{Relative Bias Ratio}

The contingency table analysis described in Section~\ref{cont_method}
tests only a very particular linear-biasing scenario (where
$b_O=b_I$).  A more general model allows for a relative bias between
the different samples (equation~\ref{bias_equ}).  An extension of the
above method allows us to test this more general model.
To allow comparison with estimates of $b_O/b_I$ from other
methods (see e.g. Section~\ref{comp_intro}), we will consider a model
where the optical over-density is a constant multiple ($b_O/b_I$) of
the IRAS over-density.

We thus construct a series of null hypotheses that the two galaxy
distributions are linearly biased with a given ratio of bias
parameters. Since we have no direct measures of the underlying mass
distribution we can only investigate the {\em ratio} of $b_O/b_I$.
For clarity we will discuss the biasing of the two populations
relative to an underlying mass distribution on the understanding that
our results are only sensitive to $b_O/b_I$.  We could test any given
ratio, perhaps arising from a physical biasing theory, or a ratio
observed from global statistical measures such as the $b/\Omega^{0.6}$
results from velocity fields.  Alternatively we could run through a
series of possible $b$ ratios and see which are the most acceptable.

To incorporate this biasing ratio we must modify our $X^2$ statistic
slightly, breaking away from a simple contingency table analysis.  For
each cell $\bar{n}_{ij} $ is estimated as $V_i\sum_i n_{ij}/\sum_i
V_i$ and $\Delta n_{ij}=n_{ij}-\bar{n}_{ij} $ ($V_i$ is the unmasked
volume of cell $i$), then
\begin{eqnarray}
\langle\frac{1}{b_j}\frac{\Delta n_{ij}}{\bar{n}_{ij} }\rangle=
\frac{\delta\rho_m}{\bar{\rho}_m}.
\end{eqnarray}
We can thus estimate the underlying mass fluctuations in a cell by
\begin{eqnarray}
\Delta_i=\frac{ \sum_j w_{ij} \Delta n_{ij}/b_j\bar{n}_{ij}  }
              { \sum_j w_{ij} }
\end{eqnarray}
We choose to use the mean densities of the two species as weights
($w_{ij}=\bar{n}_{ij} $).  This weighting scheme means our analysis
will agree with the contingency table analysis at $b_O/b_I=1$.  Other
schemes are possible but they rely on the observed values in each cell
and are not as robust.

We are then in a position to construct a new $X^2$ statistic
\begin{eqnarray}
X^2=\frac{\sum_{ij}( \Delta n_{ij}/b_j\bar{n}_{ij} -\Delta_i)^2}
            {\var( \Delta n_{ij}/b_j\bar{n}_{ij} )}.
\end{eqnarray}
The variance is determined from Poisson statistics, neglecting errors
and covariance terms arising from $\bar{n}_{ij} $ (in agreement with
our contingency table analysis if $b_O/b_I=1$):
\begin{eqnarray}
\var(\frac{1}{b_j}\frac{\Delta n_{ij}}{\bar{n}_{ij} })
=\frac{\Delta_i+1/b_j}{b_j\bar{n}_{ij} }.
\end{eqnarray}

In the previous analysis we merged cells with an estimated
under-density low enough that, for the class of galaxies with the lowest
mean density at that shell, we would expect fewer than $n_{\rm lim}$
galaxies in that cell. With our new approach a similar argument
introduces the following threshold
\begin{eqnarray}
\Delta_i>{\rm max}_j(\frac{1}{b_j}\frac{n_{\rm lim}-\bar{n}_{ij} }{\bar{n}_{ij} })
\end{eqnarray}
The results presented here have taken $n_{\rm lim}=5$.  The $\Delta_i$
limit varies with $b_O/b_I$, as does $\Delta_i$ itself so the number
of cells merged depends on the value of $b_O/b_I$ being tested.

We calculate the statistic $X^2$ for a range of $0.5<b_O/b_I<3$ and,
under the assumption that this statistic is distributed as $\chi^2$
with $M-N$ degrees of freedom, we calculate the probability
$P(X^2,\nu)$ as before.

The results are presented in Figure~\ref{bobi} which shows the
probability of obtaining a higher value of $X^2$ than that observed as
a function of the assumed biasing ratio $b_O/b_I$.  The curves in
Figure~\ref{bobi} are for the four scales 5, 10, 20 and 30$h^{-1}$
Mpc.  The curves are not smooth because our cell merging threshold
depends on the assumed bias ratio and jumps occur when new cells
become merged.  On scales of $5$ to $10h^{-1}$ Mpc {\em all} linear-biasing models relating optical and IRAS galaxies can be ruled out
with around 99 per cent significance.  At the $5h^{-1}$ Mpc scale this is
perhaps unsurprising since we are in the non-linear regime
($\delta\rho/\bar{\rho}>1$) so a naive justification for a simple
linear-biasing breaks down.  Also on these smaller scales we expect
that the `fingers of god' (more prominent for the ellipticals in
clusters) will have a non-linear impact on our analysis over and above
any discrepancies in the over-densities.  However, on scales of
$10h^{-1}$ Mpc we might have expected that a linear-biasing model
would have been reasonable.  On larger scales (20 and 30$h^{-1}$ Mpc) we do
find that there are linear-biasing models that we cannot reject.  A
bias ratio of $1.7<b_O/b_I<3.3$ cannot be rejected with more than 95 per cent
significance for 20$h^{-1}$ Mpc cells and likewise $1.1<b_O/b_I<2.0$
is acceptable at 30$h^{-1}$ Mpc.  It is interesting that we can reject
all $b_O/b_I<1.7$ with more than 95 per cent confidence since it is within
this region that most estimates of $b_O/b_I$ lie
(Section~\ref{comp_intro}).  Taken at face value this would imply that
optical and IRAS density fields are not linearly related even on
scales of 20$h^{-1}$ Mpc.

Since we were unable to reject the hypothesis that $b_O/b_I=1$ after
the exclusion of optical ellipticals, we expect a wider range of
acceptable bias ratios when comparing optical-spirals with the IRAS
sample.  This is indeed the case and for brevity we do not show the
$P(X^2,\nu)$ plots for these samples. Suffice to say we cannot rule
out $0.8<b_O/b_I<3$ between $5$ and $20h^{-1}$Mpc (all linear-biasing
models could be ruled out with 95 per cent confidence at $30h^{-1}$ Mpc but
as before the few degrees of freedom at this scale cause us to be
sceptical of this results).

\begin{figure}
\vbox to9.0in{\rule{0pt}{9.0in}}
\caption{$P(\chi^2,\nu)\; {\rm v}\; b_O/b_I$ for 
IRAS and optical galaxies on various scales.}
\label{bobi}
\end{figure}


\section{Conclusion}

We have completed two new large, deep IRAS redshift surveys.  The
large scale clustering properties of these surveys have been analysed
and found to be consistent with previous IRAS redshift surveys,
strengthening previous conclusions that the IRAS galaxy distribution
is inconsistent with the standard CDM predictions.

We noted two particular shortcomings in the counts-in-cells analysis.
Firstly, the variances determined using different grid patterns on the
same data show a scatter comparable with the quoted errors.  Secondly
the method is not completely independent of the selection function and
we demonstrate one way of overcoming this bias.

We are able to combine all the IRAS samples together to obtain a
single estimate of the variance on large scales with errors that will
not be matched until the completion of the full PSC-$z$ survey
\cite{pscz1,pscz2}.  Converting these variances into the form of a
dimensionless power-spectrum (correcting for redshift-space
distortions), we compare them with a mixed dark matter (MDM) model.  The slope of
this model spectrum appears slightly steeper than that observed but the
assumptions involved in transforming theory to data are sufficiently
uncertain that we can not read too much into this.

A comparison of our variance estimates with those from the APM-Stromlo
redshift survey suggests $b_O/b_I=1.20\pm0.05$, assuming a linear-biasing model with ratio independent of scale. This is higher than
that found by the APM-Stromlo team themselves \cite{lday_cincell}
because we find smaller IRAS variances than E90.

Using our FSS-$z$ sample and the first CfA strip we directly compare
the distributions of optical and IRAS galaxies (using a physically
meaningful definition of the two classes).  This is done on a
point-by-point basis so we are able to test the hypothesis that the
two populations are related by a linear-bias ratio ($b_O/b_I$).

We are able to reject the hypothesis that $b_O/b_I=1$ with a very high
degree of significance on the scales 5, 10 and 20 $h^{-1}$ Mpc.  This
hypothesis is, however, acceptable if we exclude ellipticals and S0
galaxies from the optical sample, suggesting that the discrepancies
are due to the morphology density relation and the
under-representation of ellipticals in IRAS samples.  On scales of 30
$h^{-1}$ Mpc our conclusions would be reversed but we suspect that our
statistics are too poor at this scale for this to be very meaningful.

Allowing $b_O/b_I$ to vary away from 1 we find that only on the
largest scales are {\em any} optical/IRAS linear-biasing models
acceptable, if elliptical galaxies are included in the optical
samples.  Even on scales of 20 $h^{-1}$ Mpc, linear-biasing with $b_O/b_1<1.7$
is ruled out with more than 95 per cent confidence.  It should of course be
stressed that in this particular volume the optical galaxy
distribution is dominated by the Coma cluster and so the distorting
effect of the elliptical galaxies has to some extent been amplified.

This non-linear clustering relation between ellipticals and IRAS
galaxies may present complications for the interpretation of any
galaxy surveys that include both elliptical and spiral galaxies.  As
yet we have been unable to determine any differences in the clustering
properties of IRAS galaxies and spiral galaxies and so surveys
composed of either of these can be meaningfully compared.

\vspace{1cm}

{\bf ACKNOWLEDGMENTS}

SJO and ANT acknowledge PPARC support.  WS was supported by a PPARC
Advanced Fellowship.  We thank John Huchra, Ray
Wolstencroft, Quentin Parker \& Roger Clowes and Marc Davis \& Michael
Strauss for providing redshifts in advance of publication.  We
acknowledge use of the ING telescopes and thank the staff for their
invaluable assistance.  Much use has been made of the STARLINK
resources.

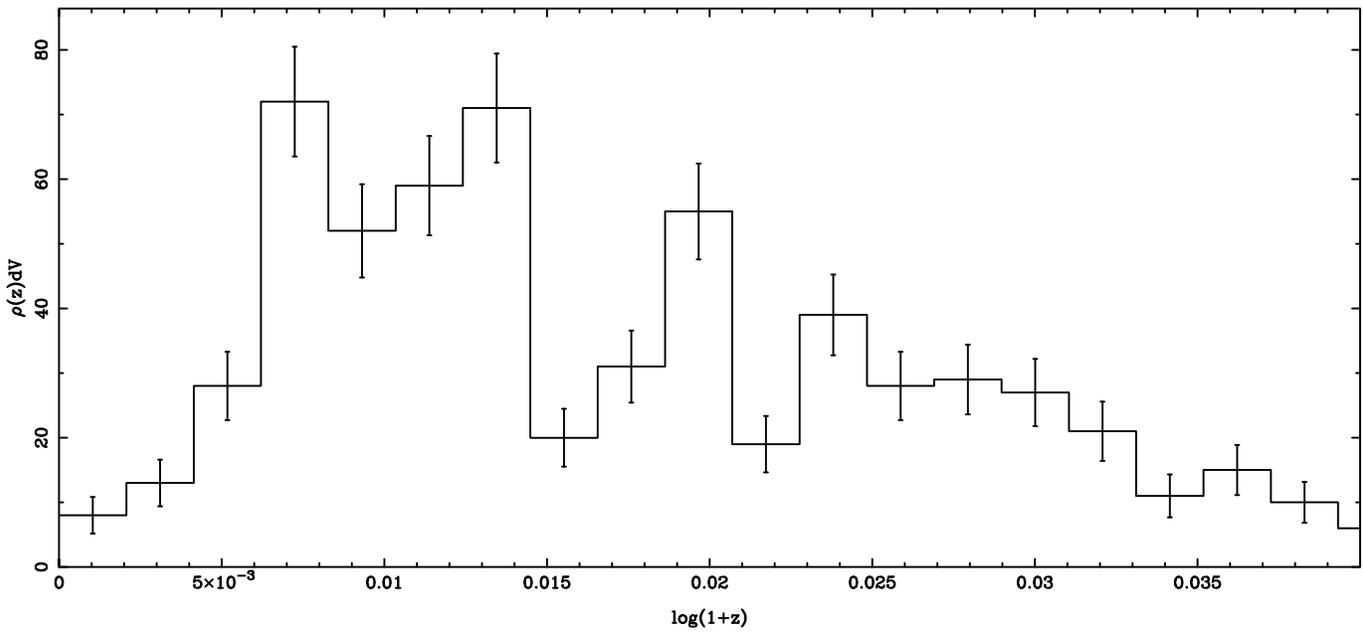
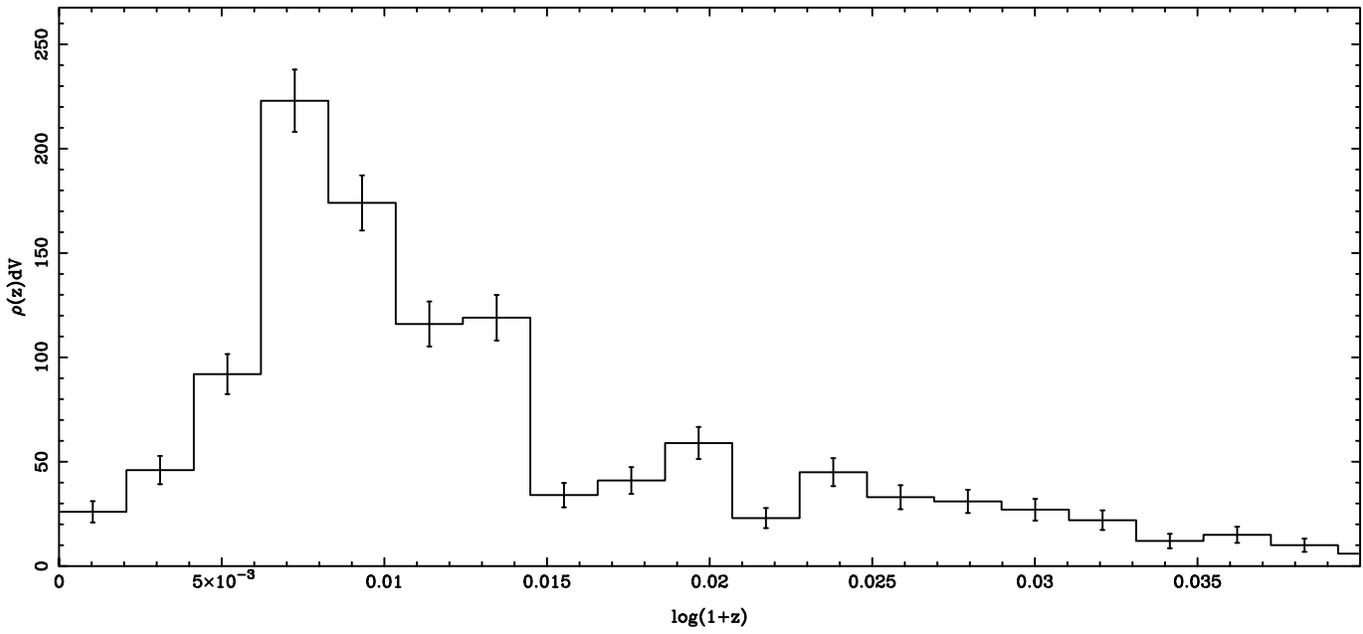
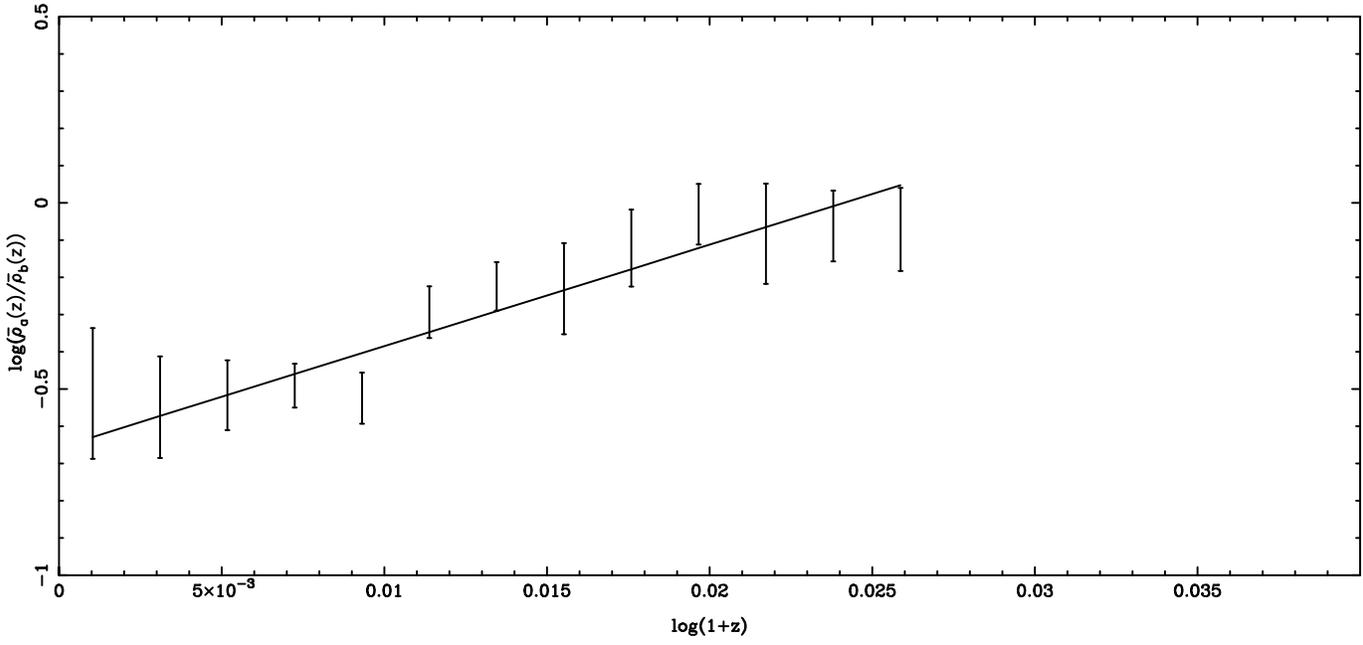
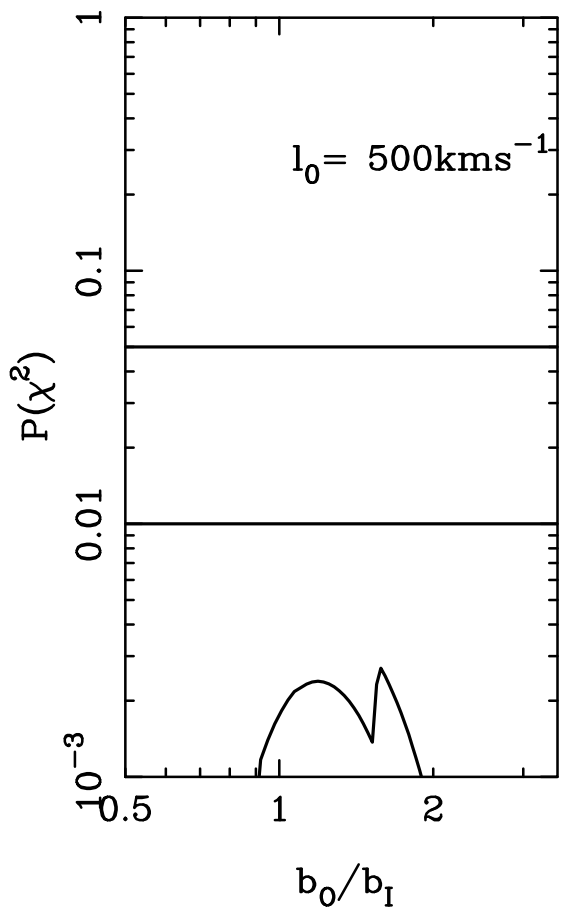
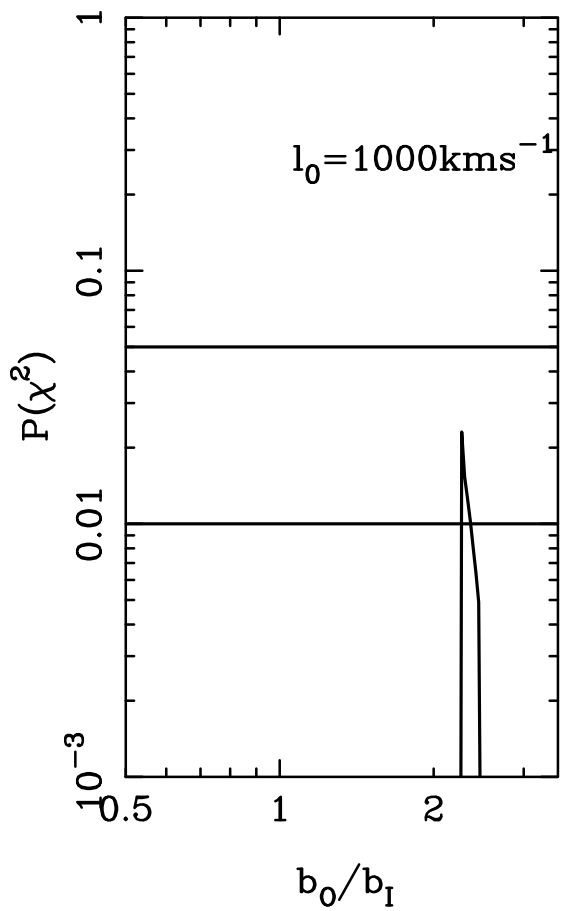
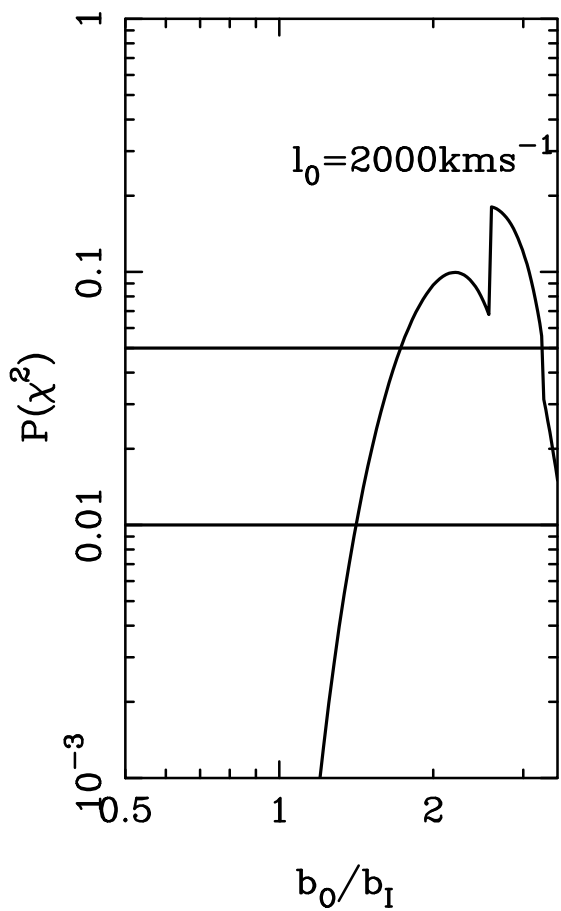
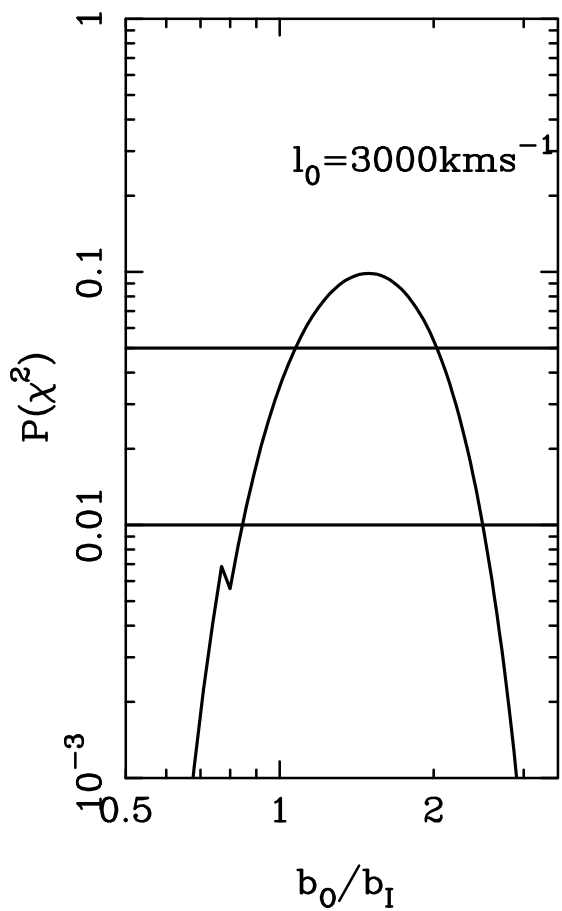
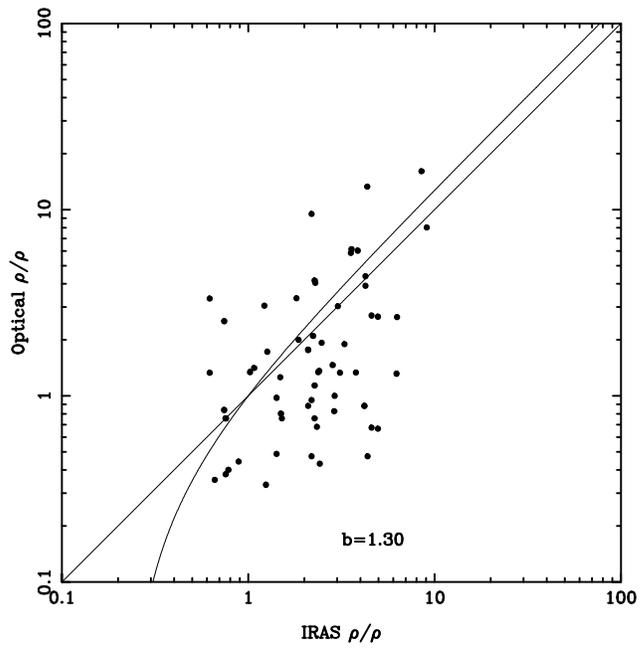
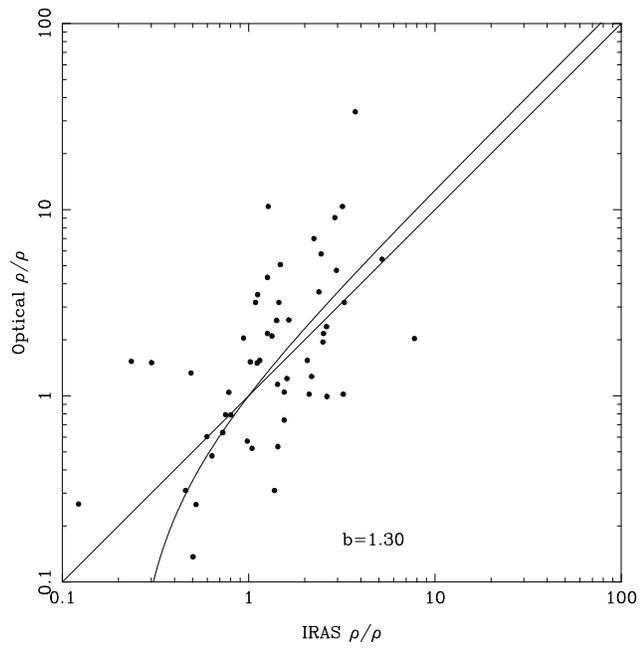
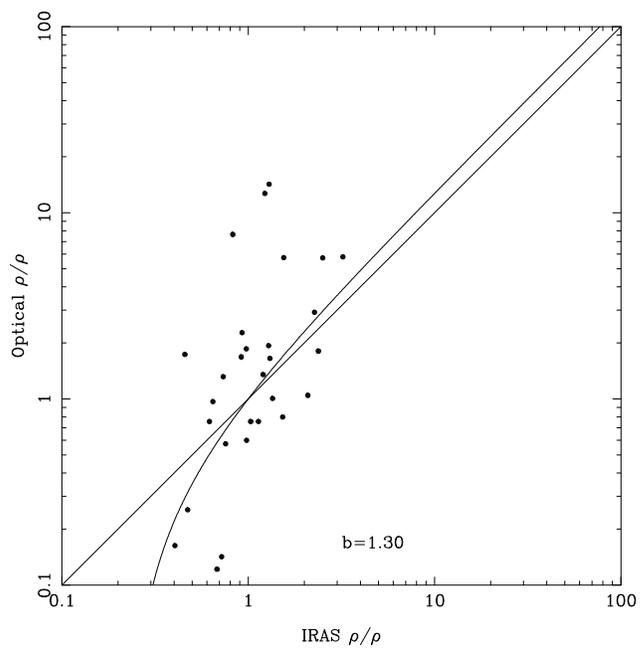
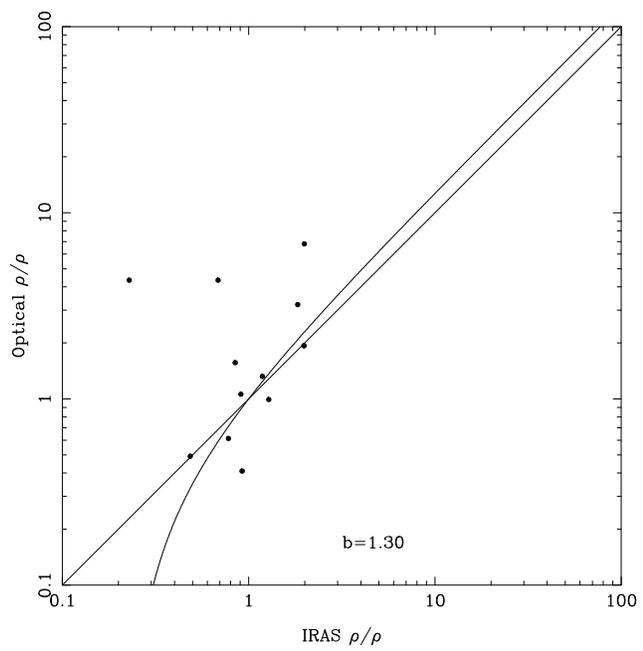
\end{document}